\documentclass[prl,twocolumn,showpacs,preprintnumbers,amsmath,amssymb,superscriptaddress]{revtex4}
\usepackage{graphicx}
\begin{document}

\title{Ultra-sensitive surface absorption spectroscopy using
sub-wavelength diameter optical fibers}
\author{F. Warken}

\affiliation{Institut f\"ur Angewandte Physik, Universit\"at Bonn,
53012 Bonn, Germany}

\author{E. Vetsch}

\affiliation{Institut f\"ur Angewandte Physik, Universit\"at Bonn,
53012 Bonn, Germany}

\author{D. Meschede}

\affiliation{Institut f\"ur Angewandte Physik, Universit\"at Bonn,
53012 Bonn, Germany}

\author{M. Sokolowski}

\affiliation{Institut f\"ur Physikalische und Theoretische Chemie,
Universit\"at Bonn, 53012 Bonn, Germany}

\author{A. Rauschenbeutel}%

\altaffiliation{Present address: Institut f\"ur Physik,
Universit\"at Mainz, 55099 Mainz, Germany}

\email{rauschenbeutel@uni-mainz.de}

\affiliation{Institut f\"ur Angewandte Physik, Universit\"at Bonn,
53012 Bonn, Germany}

\date{\today}

\begin{abstract}

The guided modes of sub-wavelength diameter air-clad optical fibers
exhibit a pronounced evanescent field. The absorption of particles
on the fiber surface is therefore readily detected via the fiber
transmission. We show that the resulting absorption for a given
surface coverage can be orders of magnitude higher than for
conventional surface spectroscopy. As a demonstration, we present
measurements on sub-monolayers of 3,4,9,10-perylene-tetracarboxylic
dianhydride (PTCDA) molecules at ambient conditions, revealing the
agglomeration dynamics on a second to minutes timescale.
\end{abstract}


\maketitle

During the last twenty years, numerous optical tools for surface
and interface analysis have been developed \cite{Bor05}. The
selective sensitivity to surface effects is often obtained by
carrying out spectroscopy with evanescent waves (EW), created by
total internal reflection of light at the interface. This is
straightforwardly realized by exciting waveguide modes in unclad
optical fibers \cite{Paul87,Messica96}. If the EW is resonant with
the transition frequency of particles (atoms, molecules, quantum
dots, etc.) in the surrounding medium, one can use both the
particles' fluorescence \cite{Fang99} or the peak attenuation of
the wave\-guide mode \cite{Paul87,Simhony88} to infer the
concentration of particles at the interface. Moreover, the line
shapes allow to spectroscopically retrieve detailed physical
information about the nature and strength of the particle--surface
interaction.

Fiber-based evanescent wave spectroscopy (EWS) is used in various
sensors \cite{Potyrailo98}. The robustness, reliability, and ease
of use of an all-fiber-based sensor technology is advantageous for
in situ sensing in a remote or isolated location or in a harsh
environment, e.g., in industrial applications or environmental
studies. Furthermore, such sensors also profit from the
multiplexing and miniaturization potential inherent to fiber
technology. When measuring a volumetric concentration of particles
in the surrounding medium, these sensors yield however a reduced
sensitivity compared to conventional free-beam absorption: a
significant fraction of the light propagates inside the waveguide
and therefore does not interact with the particles of interest.
This problem can partially be overcome by increasing the power
fraction in the EW through proper choice of the fiber mode or
geometry \cite{Gupta96,Tai87,Lou05}. Yet, even in the ultimate case
of 100~\% EW, the sensitivity will not exceed that of free-beam
absorption techniques.

\begin{figure}
\includegraphics[width=6.9cm]{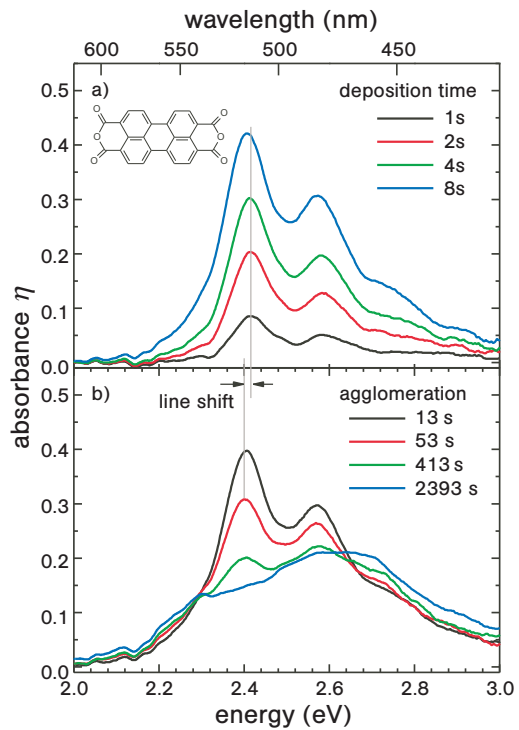}
\caption{\label{fig:Aufdampfdynamik} (Color online) Consecutive
absorbance spectra. (a) Deposition of a sub-monolayer: more and
more molecules still show a monomer-like spectrum. (b) Evolution of
the spectral absorption of a constant molecule number. The shape
varies continuously from monomer-like to oligomer-like. Thus we
observe agglomeration of the molecules on the fiber surface.}
\end{figure}

In this letter, we demonstrate that the situation can be
dramatically different when employing fiber-based EWS for the
spectroscopic study of surface coverages instead of volumetric
concentrations: The ultimate sensitivity of fiber-based surface
absorption spectroscopy (SAS) is shown to strongly depend on the
fiber diameter and to exceed free-beam SAS by several orders of
magnitude in the case of sub-wavelength diameter fibers.
Fiber-based surface absorption spectroscopy (SAS) has already been
used for a number of applications, e.g., in
bio-sensors~\cite{Marazuela02}. However, to our knowledge, these
applications were only motivated by the above-mentioned practical
advantages; their ultimate sensitivity was so far neither the
subject of theoretical nor of experimental studies. As a
demonstration, we present fiber-based SAS measurements on
sub-monolayers of 3,4,9,10-perylene-tetracarboxylic dianhydride
molecules (PTCDA, see inset of Fig.~\ref{fig:Aufdampfdynamik}(a)
below). PTCDA thin films are excellent indicators for the
sensitivity because their spectral properties highly depend on the
detailed arrangement of the molecules on the surface
\cite{Proehl04}. Our findings are of relevance for a large variety
of fields ranging from sensorial applications in industry,
environmental studies, and bio-technology to fundamental research
concerning thin film growth as well as the controlled interaction
of light and matter at the ultimate, microscopic scale.

We start by considering the absorbance (i.e., decadal absorption
coefficient) of a dilute film of molecules deposited on a
dielectric surface. It is defined as
\begin{equation}\label{eq:def_absorbance}
    \eta(\lambda)=
    -\lg\left(\frac{P_\mathrm{sig}(\lambda)}
    {P_\mathrm{ref}(\lambda)}\right)\approx
    \frac{P_\mathrm{abs}(\lambda)}
    {\ln (10) P_\mathrm{ref}(\lambda)}
    \ ,
\end{equation}
where $P_\mathrm{sig}(\lambda)$ and $P_\mathrm{ref}(\lambda)$ are
the transmitted powers within the spectral interval
$[\lambda,\lambda+ \Delta\lambda]$ with and without molecules
present, respectively. The right hand side of
Eq.~(\ref{eq:def_absorbance}) holds in the limit of a weak
absorption, i.e., for $P_\mathrm{abs}(\lambda)=
P_\mathrm{ref}(\lambda)-P_\mathrm{sig}(\lambda)\ll
P_\mathrm{ref}(\lambda)$.

The absorbance of a freely propagating light beam impinging onto
the film is proportional to the surface coverage $\theta$, i.e.,
the number of molecules $n$ per surface area:
\begin{equation}\label{eq:free_beam_absorption_1}
    \eta_\mathrm{free}(\lambda)=
    \theta\sigma(\lambda)/\ln (10) \ ,
\end{equation}
where $\sigma(\lambda)$ is the molecules' absorption cross section.
By choosing a non-zero angle of incidence $\phi$,
$\eta_\mathrm{free}$ can in principle be enhanced by a factor of
$1/\cos(\phi)$. However, this factor is limited by the experimental
geometry and typically remains of order one. For free beam
measurement schemes (e.g., transmission spectroscopy, differential
reflection spectroscopy (DRS), or attenuated total reflection
spectroscopy, see \cite{Bor05} for a review), the smallest
detectable surface coverage is thus determined by the smallest
detectable absorbance according to
\begin{equation}\label{eq:free_beam_absorption_2}
    \theta_\mathrm{min}\approx \ln (10)
    \eta_\mathrm{min}(\lambda)/\sigma(\lambda)\ .
\end{equation}

The situation is very different in fiber-based SAS, where every
molecule absorbs a constant fraction
$\sigma(\lambda)/A_\mathrm{eff}$ of the power with $A_\mathrm{eff}$
given by
\begin{equation}\label{eq:A_eff}
    A_\mathrm{eff}(\lambda)=
    P_\mathrm{ref}(\lambda)/
    I_\mathrm{surf}(\lambda)\ .
\end{equation}
$A_\mathrm{eff}$ thus corresponds to an effective area of the
guided fiber mode which is normalized to the evanescent field
intensity at the fiber surface, $I_\mathrm{surf}(\lambda)$,
calculated according to \cite{Le_Kien06}. The transmitted power
after interaction with $n$ molecules is then given by
\begin{equation}\label{eq:fibre_absorbed_power}
    P_\mathrm{sig}(\lambda)=
    P_\mathrm{ref}(\lambda)
    \left[1-\sigma(\lambda)/A_\mathrm{eff}\right]^n\ ,
\end{equation}
yielding an absorbance of
\begin{equation}\label{eq:fibre_absorbance}
    \eta_\mathrm{fiber}(\lambda)=
    -\lg\left(\frac{P_\mathrm{sig}(\lambda)}
    {P_\mathrm{ref}(\lambda)}\right)\approx
    \frac{n\sigma(\lambda)}{\ln (10)
    A_\mathrm{eff}(\lambda)}\ ,
\end{equation}
where the right hand side holds for
$\sigma(\lambda)/A_\mathrm{eff}\ll 1$. Substituting $n=2\pi
RL\theta$, where $R$ and $L$ are the fiber radius and length,
respectively, then yields
\begin{equation}\label{eq:fibre_absorbance_1}
    \eta_\mathrm{fiber}(\lambda)\approx
    \frac{\theta\sigma(\lambda)}{\ln (10)}\cdot
    \frac{2\pi RL}{A_\mathrm{eff}(\lambda)}=
    \eta_\mathrm{free}(\lambda) \xi(\lambda)
    \ .
\end{equation}
For a given surface coverage $\theta$, the absorbance in
fiber-based SAS is thus enhanced by a factor of $\xi(\lambda)=2\pi
RL/A_\mathrm{eff}(\lambda)$. As a first estimate, we can assume
$A_\mathrm{eff}(\lambda)\approx \pi R^2$, yielding $\xi\approx
2L/R$. With $L$ of the order of millimetres and $R$ in the
sub-micron range, we thus expect $\xi$ to be as large as 10000 for
typical ultra-thin fibers, promising an increase of four orders of
magnitude in sensitivity.
\begin{figure}
\includegraphics[width=8cm]{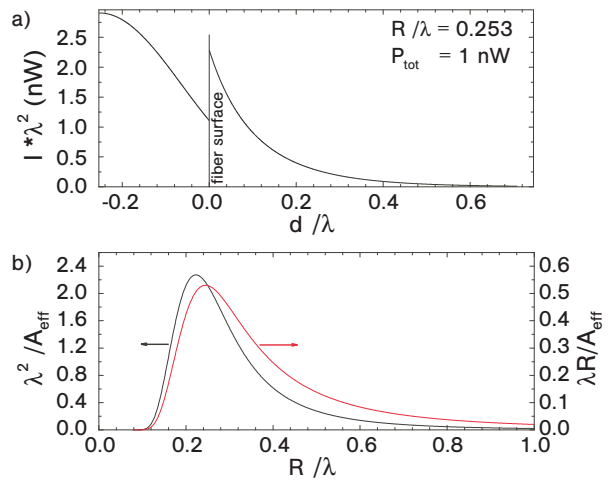}
\caption{\label{fig:1_over_A_eff} (Color online) (a) Intensity
profile of the fundamental guided HE$_{11}$ mode of an air-clad
silica fiber in units of Nanowatts per $\lambda^2$ as a function of
the distance $d$ from the fiber surface in units of $\lambda$.
Calculated according to \cite{Le_Kien06} for unpolarized light, a
total guided power of 1~nW, and a fiber radius of
$0.253\times\lambda$. (b) Plot of $1/A_\mathrm{eff}$ and
$R/A_\mathrm{eff}$ in units of $1/\lambda^2$ and $1/\lambda$,
respectively, as a function of the fiber radius $R$ in units of
$\lambda$. Assuming a refractive index of 1.46, all plots in (a)
and (b) hold universally for any $\lambda$. }
\end{figure}
For a more rigorous estimate, we calculate
$A_\mathrm{eff}(\lambda)$ from Eq.~(\ref{eq:A_eff}) using the
intensity profile of the fundamental guided mode of the air-clad
silica fiber, see Fig.~\ref{fig:1_over_A_eff}(a).
Figure~\ref{fig:1_over_A_eff}(b) gives $R/A_\mathrm{eff}(\lambda)$
in units of $1/\lambda$ as a function of $R$ in units of $\lambda$
(red solid line). Interestingly, $R\lambda/A_\mathrm{eff}(\lambda)$
reaches a maximum for $R/\lambda=0.253$ and then rapidly goes to
zero because for $R\ll \lambda$ the mode is only weakly bound and
$A_\mathrm{eff}$ diverges. When seeking to detect or to
characterize a given surface coverage of molecules with the highest
possible sensitivity, one should thus choose a fiber radius of
approximately one fourth of the central wavelength
$\lambda_\mathrm{c}$ of the measured spectrum. The enhancement
factor with respect to free beam spectroscopy in this case will be
$\xi\approx 3.64\times L/\lambda_\mathrm{c}$. For
$\lambda_\mathrm{c}=500$~nm, $\xi\approx 10000$ is, e.g., already
obtained for $L=1.5$~mm, which can easily be realized for
ultra-thin optical fibers.

We note that, due to the wavelength dependence of $\xi(\lambda)$,
the measured absorbance $\eta_\mathrm{fiber}(\lambda)$ is not
directly proportional to $\sigma(\lambda)$ as in the case of free
beam spectroscopy. Before interpretation, the spectra will thus
have to be normalized with respect to $\xi(\lambda)$, as calculated
from Fig.~\ref{fig:1_over_A_eff}(b). Finally, according to
Eq.~(\ref{eq:fibre_absorbance}), if one seeks to detect or to
characterize a given number of particles on the fiber surface
rather than a given surface coverage, one will have to maximize
$1/A_\mathrm{eff}(\lambda)$ instead of $R/A_\mathrm{eff}(\lambda)$.
This is realized for a 10~\% smaller fiber radius of
$R/\lambda=0.228$, see black solid line in
Fig.~\ref{fig:1_over_A_eff}(b).

In order to experimentally illustrate the enhanced sensitivity of
the described method with respect to free beam measurement schemes,
we have deposited PTCDA molecules on the surface of an ultra-thin
fiber waist. PTCDA is an interesting model system for organic thin
film growth. It forms smooth and highly ordered layers on a large
variety of different substrates \cite{Forrest97} and shows clear
spectroscopic signatures for different phases \cite{Proehl04}.

We fabricate the tapered fibers by stretching a standard optical
single mode fiber (Newport F-SF) while heating it with a travelling
hydrogen/oxygen flame \cite{Birks92}. Our computer controlled fiber
pulling rig produces tapered fibers with a homogeneous waist
diameter down to 100~nm and a typical extension of 1--10~mm. By
analytically modelling the pulling process, we can predict the
resulting fiber profile with better than 10\% precision. In the
taper sections, the weakly guided LP$_{01}$ mode of the unstretched
fiber is adiabatically transformed into the strongly guided
HE$_{11}$ mode of the ultrathin section and back \cite{Love86},
resulting in a highly efficient coupling of light in and out of the
taper waist. For monochromatic light of 850-nm wavelength and
fibers with a final diameter above 0.5~$\mu$m, we reach up to 97~\%
of the initial transmission.

\begin{figure}
\includegraphics[width=8cm]{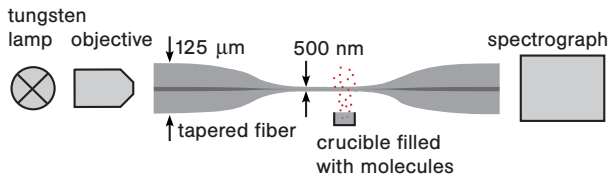}
\caption{\label{fig:setup} (Color online) Scheme of the
experimental set-up. White light from a tungsten lamp is
transmitted through a tapered fiber with a 500-nm diameter waist
and analyzed by a CCD spectrograph. This allows to measure the
absorbance of molecules deposited on the fiber waist with a high
sensitivity.}
\end{figure}

A so-prepared 500-nm diameter tapered fiber with a waist length of
3~mm was used for our measurements. Figure~\ref{fig:setup} shows
the simple experimental set-up, using a conventional absorption
spectrometer configuration with a tungsten white light source and a
commercial CCD spectrograph (Ocean Optics SW1024). A reference
spectrum with 6~nm effective spectral resolution is recorded
without molecules.

The molecules are deposited on the fiber waist by placing a
crucible with PTCDA crystals below the fiber and by heating it up
to $250^\circ \mathrm{C}$. By convection, the air carries
sublimated molecules to the fiber waist where they are adsorbed.
During deposition, we continuously record spectra with an
integration time of 1~s.
Figure~\ref{fig:Aufdampfdynamik}(a) displays a series of absorption
spectra, recorded with about 25~nW of white light, not saturating
the molecules. All spectra show a clear vibronic progression.
Qualitatively, they agree well with DRS spectra of sub-monolayers
of PTCDA on mica reported in \cite{Proehl04}. Quantitatively, the
absorbance on the high energy side is reduced by 5--10~\% in our
case due to the wavelength-dependance of the enhancement factor
$\xi$. From the absorbance of 0.09--0.42 at 2.4~eV and using
$\sigma=2.7\times 10^{-16}~\mathrm{cm}^{2}$ \cite{Hoff00} in
Eq.~(\ref{eq:fibre_absorbance_1}), we infer that 0.5--$2.3\times
10^7$ molecules covered the fiber waist. This corresponds to a
surface coverage of 1.0--$4.9 \times 10^{11}$~cm$^{-2}$ or
0.12--0.59~\% of a compact monolayer of flat lying PTCDA molecules,
arranged in the herringbone structure of the (102) plane of the
PTCDA bulk crystal \cite{Forrest94}. Thus, with respect to the DRS
spectra reported in \cite{Proehl04}, the smallest coverages in our
case were about two orders of magnitude smaller. The strongest line
at 2.42~eV (see the 1--4~s spectra) shifts by 0.02~eV to smaller
energies between 4 and 53 s. This shift has been observed
previously and is attributed to the condensation of isolated PTCDA
molecules into two-dimensional islands \cite{Proehl04}. This
clearly demonstrates that sub-monolayer absorbance spectra can be
measured very rapidly and with an excellent signal to noise ratio
with fiber-based SAS.

Figure~\ref{fig:Aufdampfdynamik}(b) displays a series of spectra
that monitor the post-deposition evolution, i.e., the ripening of
the film after stopping the deposition of the molecules. We observe
a continuous transformation from a monolayer to a multilayer
spectrum, similar to \cite{Proehl05}. It demonstrates that a PTCDA
monolayer on a glass surface is meta-stable at ambient conditions
and transforms into islands with a thickness of at least two
monolayers within minutes. Two isosbestic points are observed at
2.30~eV and 2.67~eV, confirming that the total number of molecules
is constant during the ripening.

\begin{figure}
\includegraphics[width=7cm]{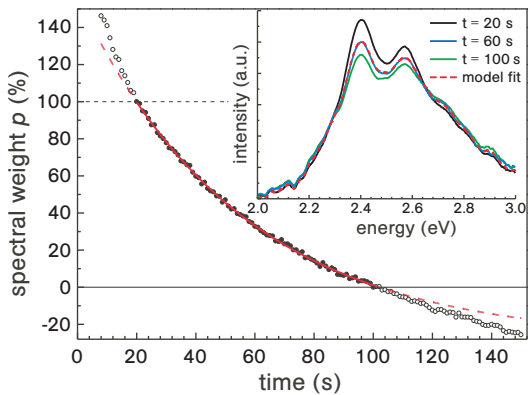}
\caption{\label{fig:Model} (Color online) Spectral investigation of
the film ripening. Spectra for $t=20$--100~s can be modelled as a
weighted sum of the spectra at $t=20$~s and $t=100$~s. As an
example, we show the $t=60$~s spectrum and the corresponding model
fit (see inset). The spectral weight of the $t=20$~s spectrum
decays according to an offset exponential for $t=20$--100~s
\cite{Offset}. The corresponding fit yields a decay constant of
$55(\pm 1)$~s. The dotted line extrapolates the fit for $t<20$~s
and $t>100$~s. }
\end{figure}

Due to the fast data collection, we are able to extract additional
kinetic details of the ripening process. In the time interval
$t=20$--100~s, our spectra can be understood as a weighted sum,
$p\cdot A+(1-p)\cdot B$, of a monolayer spectrum $A$ (the 20~s
spectrum in the inset of Fig.~\ref{fig:Model}) and a spectrum $B$,
where most PTCDA molecules have a next neighbour vertical to the
glass surface and are thus dimers concerning their optical
properties (the 100~s spectrum in Fig.~\ref{fig:Model}). We fitted
this model to our data. As an example, the 60~s spectrum and its
fit are shown. The weight of the monolayer spectrum is plotted as a
function of time in Fig.~\ref{fig:Model}. For $t=20$--100~s it
decays with a time constant of $55(\pm 1)$~s according to an offset
exponential \cite{Offset}, indicating first order kinetics of the
ripening. We thus conclude that the molecules reorganize into
dimers through a growth process rather than through two-body
collisions, because the latter would lead to second order kinetics.
Outside of the $t=20$--100~s interval, the experimental data
deviate from the extrapolated exponential behaviour (dotted line),
indicating that channels other than $A\rightarrow B$ significantly
contribute to the transformation dynamics. We attribute this to the
above-mentioned condensation of isolated PTCDA molecules into
two-dimensional islands for $t<20$~s and to the formation of
oligomers for $t>100$~s.

Reordering of monolayer islands to a polycrystalline phase was also
observed for PTCDA deposited on mica when transferring the samples
from vacuum to ambient conditions \cite{Proehl05}. The
agglomeration process was however not resolved in time. By
comparison with samples in a dry atmosphere, Proehl {\it et
al.}~deduced that coadsorbed water accelerates the reordering
\cite{Proehl05}. The small time constant observed in our
experiment, entirely performed under ambient conditions, confirms
this conclusion. Moreover, our experiment demonstrates that the
metastable monolayer exists only for a very short time under
ambient conditions and can hence be easily overlooked.

Summarizing, we have shown that sub-wavelength diameter optical
fibers are a powerful and easy to use tool for ultra-sensitive
surface absorption spectroscopy. In a theoretical analysis we
showed that the sensitivity of fiber-based surface absorption
spectroscopy can outperform free-beam techniques by several orders
of magnitude. We applied our method to sub-monolayers of PTCDA
molecules, deposited on the surface of an ultra-thin fiber. The
high sensitivity allowed us to record spectra with an excellent
signal to noise ratio in a short time, revealing the dynamics of
the deposition and of the post-deposition evolution of the dilute
molecular film on a second to minutes time scale. The experiments
where carried out with an extremely simple and low-cost
experimental set-up. Yet, the sensitivity of our measurements
exceeded what was reached in previous studies by two orders of
magnitude. Straightforward technical improvements should allow to
further increase this value and to approach the ultimate
sensitivity of the technique.

We wish to thank H.~Giessen and Y.~Louyer for their support in the
early stages of the experiment and W.~Alt, T.~Fritz, and
M.~Hoffmann for valuable discussions. This work was supported by
the EU (Research Training Network ``FASTNet'') and the DFG
(Research Unit 557).


\begin{thebibliography}{99}
\bibitem{Offset}
The offset accounts for the fact that spectrum $B$
%corrsponds to an intermediate spectrum which
still contains a monolayer component and thus an admixture of
spectrum $A$.
\end{thebibliography}
\end{document}